\documentclass[lettersize,journal]{IEEEtran}
\usepackage[utf8]{inputenc}
\usepackage[T1]{fontenc}
\usepackage{amsmath,amsfonts}
\usepackage{algorithmic}
\usepackage{array}
\usepackage[caption=false,font=normalsize,labelfont=sf,textfont=sf]{subfig}
\usepackage{textcomp}
\usepackage{stfloats}
\usepackage{url}
\usepackage{verbatim}
\usepackage{graphicx}
\usepackage{colortbl}
\usepackage{makecell}
\usepackage{tabularx}
\usepackage{multirow}
\newcolumntype{Y}{>{\centering\arraybackslash}X}
\usepackage{hyperref}
\hypersetup{
	colorlinks=true,
	linkcolor=blue,
	filecolor=magenta,      
	urlcolor=blue,
	citecolor=blue
}

\usepackage[nolist]{acronym}

\def\BibTeX{{\rm B\kern-.05em{\sc i\kern-.025em b}\kern-.08em
    T\kern-.1667em\lower.7ex\hbox{E}\kern-.125emX}}
\usepackage{balance}

\usepackage{multirow}
\usepackage{booktabs}
\usepackage{graphics}
\usepackage[dvipsnames,table,xcdraw]{xcolor}

\begin{document}

\acrodef{ML}{Machine Learning}
\acrodef{AI}{Artificial Intelligence}
\acrodef{XAI}{eXplainable Artificial Intelligence}
\acrodef{MLP}{Multilayer Perceptron}
\acrodef{XGB}{eXtreme Gradient Boosting}
\acrodef{SVM}{Support Vector Machine}
\acrodef{RL}{Reinforcement Learning}
\acrodef{QoT}{Quality of Transmission}
\acrodef{SHAP}{SHapley Additive Explanations \cite{lundberg2017unified}}
\acrodef{FS}{Feature Selection}
\acrodef{OPM}{Optical Performance Monitoring}
\acrodef{BAC}{Balanced Accuracy}
\acrodef{F1}{F1-score}
\acrodef{G-Mean}{Geometric Mean}

\title{Explainable and Resilient ML-Based Physical-Layer Attack Detectors}

\author{Aleksandra Knapi{\'n}ska,~\IEEEmembership{Member,~IEEE} and Marija Furdek,~\IEEEmembership{Senior~Member,~IEEE}
\thanks{Authors are with the Department of Electrical Engineering, Chalmers University of Technology, Gothenburg, Sweden (e-mails: \{alekna, furdek\}@chalmers.se). Aleksandra Knapi{\'n}ska is also with the Department of Systems and Computer Networks, Wroc\l{}aw University of Science and Technology, Wroc\l{}aw, Poland. This work was supported by the Swedish Research Council (2023–05249) and the European Commission’s Digital Europe Programme (101127973) through the 5G-TACTIC project.}
\thanks{Manuscript received September 30, 2025.}}



\maketitle

\begin{abstract}
Detection of emerging attacks on network infrastructure is a critical aspect of security management. To meet the growing scale and complexity of modern threats, machine learning (ML) techniques offer valuable tools for automating the detection of malicious activities. However, as these techniques become more complex, their internal operations grow increasingly opaque.
In this context, we address the need for explainable physical-layer attack detection methods. First, we analyze the inner workings of various classifiers trained to alert about physical layer intrusions, examining how the influence of different monitored parameters varies depending on the type of attack being detected. This analysis not only improves the interpretability of the models but also suggests ways to enhance their design for increased speed. In the second part, we evaluate the detectors’ resilience to malicious parameter noising. The results highlight a key trade-off between model speed and resilience. This work serves as a~design guideline for developing fast and robust detectors trained on available network monitoring data. 
\end{abstract}


\section{Introduction}
\IEEEPARstart{N}{etwork} attack detection is crucial for effective network security management, particularly when protecting critical, fiber-based communication network infrastructure \cite{furdek2021optical}. Optical networks underpin global communications: they connect different regions and continents through the long-haul, core network segment, interconnect cities through the metro segment, and provide broadband connectivity to end users (e.g., base stations, businesses, and households) through the access segment. 
Their integral role highlights the importance of optical network security. The physical layer is vulnerable to various malicious activities aimed at service disruption, ranging from deliberate fiber cuts to more subtle techniques such as insertion of jamming signals or polarization scrambling. Security breaches at the optical layer can have rippling disruptive effects across multiple tiers stacked on top of it and can jeopardize the upper-layer services regardless of the application-level security mechanisms. Hence, physical-layer attack detection is key for quick threat remediation and service down time reduction.
 
As attacks become more sophisticated, they often manifest as only minor deviations in telemetry data collected by mo\-ni\-to\-ring systems, making detection increasingly challenging. \ac{ML}-based methods offer a promising solution, as they can detect subtle anomalies in individual measurements and trigger alerts upon identifying suspicious patterns.
However, the high complexity and black-box nature of these algorithms make them less trustworthy and, therefore, difficult to deploy in modern networks. Understandably, ope\-ra\-tors are hesitant to rely on solutions whose internal workings and decision-making processes remain opaque. Furthermore, imprecise definition of objective function and other human errors might sometimes lead to hazardous decisions taken by the employed \ac{AI} agents \cite{long2024safe}. In this context, \ac{XAI} tools provide a means to improve the interpretability of complex models \cite{adadi2018peeking}. Contemporary literature investigates two main approaches to achieving model transparency. The first involves using inherently interpretable models, such as decision trees. While these offer full transparency, they often do so at the cost of sub-optimal performance. The second approach involves employing high-performing models and explaining their behavior in a \emph{post hoc} manner, i.e., after training. This method preserves model performance but provides only indirect insight into the model’s decision-making process \cite{lundberg2017unified}. 

Beyond interpretability, another critical concern when deploying \ac{ML}-based solutions for network infrastructure security is the security of the models themselves. Even well-performing and explainable models can be vulnerable to malicious interventions, such as noising attacks or adversarial manipulations, which can degrade their reliability or cause them to fail silently \cite{kumar2020adversarial}. Like any automated component in a critical system, \ac{ML} models must be evaluated not only for their accuracy and transparency but also for their resilience to such threats. 

In this paper, we address the need for developing explainable physical-layer attack detection techniques, resilient to parameter noising.
To this end, we first investigate the performance of three distinct classifiers for detecting attacks at the network's physical layer. Using a real-world dataset collected in a laboratory environment, we analyze how feature importance varies depending on the type of attack and the classifier used. Additionally, we examine the classifiers’ ability to generalize and detect attacks in aggregated data scenarios. Then, we leverage the insights gained to optimize the classifiers and reduce the amount of training data required for effective detection. Finally, we test the resilience of the classifiers before and after their optimization through a parameter noising experiment. The conducted analysis reveals crucial tradeoffs between detection speed and model resilience. 

The remainder of the paper is organized as follows: Section~\ref{sec:related_work} reviews recent advances in research related to various aspects of this work. Section~\ref{sec:experimental_environment} describes the experimental environment and dataset used in this study. Section~\ref{sec:xai_analysis} provides a detailed analysis of the created attack detectors using \ac{XAI} tools. Section~\ref{sec:detector_optimization} discusses \ac{XAI}-based detector optimization and its detailed evaluation. Section~\ref{sec:detector_resilience} presents the conducted detector resilience experiments and their implications. Finally, Section~\ref{sec:conclusions} concludes the paper. 

\section{Related Work}
\label{sec:related_work}

The security of the physical layer in ever-evolving networks is becoming increasingly critical due to the growing volume and value of transmitted data. As a result, there is a rising demand for security management schemes, particularly in relation to the crucial issue of attack detection \cite{skorin2016physical}. \ac{ML} techniques, capable of identifying subtle changes in \ac{OPM} parameters, have proven to be particularly effective tools for detecting various intrusions \cite{furdek2020machine,natalino2022rootcause}. 

There are numerous classical attacks that recur over time, providing operators with a substantial amount of historical data on how they manifest. Using this data, specialized detectors based on supervised learning can be trained to raise an alarm if such intrusions happen again \cite{sakhnini2021physical,hoang2021physical}. On the other hand, malicious actors continuously refine their techniques to invent new, previously unseen attack types, which cannot be easily detected by specialized models. In this context, unsupervised learning techniques excel at detecting anomalies in telemetry data, alerting operators to suspicious behaviors occurring within their networks \cite{natalino2022microservice,lechowicz2024trade}. Finally, binary classifiers trained on combined data from various attack types offer a promising solution to the problem of ever-evolving attack techniques, allowing models to continuously improve \cite{knapinska2025dynamic,knapinska2025experimental}. 

However, \ac{ML}-based models, which operate as black boxes, are not easily interpretable, and their decisions can be difficult to explain to operators, especially if they are unexpected. As a result, \ac{XAI} techniques are being adopted to enhance the understanding and trust in these models. \ac{QoT} estimation serves as a prime example of a~network-related field where \ac{XAI} has enabled the interpretation, optimization, and validation of models in terms of their speed, accuracy, and uncertainty \cite{ayoub2022towards,ayoub2022application,ayoub2022quantifying,houssiany2023using}. Furthermore, efforts to make these explanations more accessible are underway by providing interpretability information in natural language \cite{ayoub2025natural}. 

Other areas of network-related research where \ac{XAI} has recently enabled substantial improvements include network traffic analysis and prediction \cite{morichetta2019explain,knapinska2024explainable,knapinska2025explaining}, as well as \ac{ML}-assisted resource allocation \cite{goscien2025explainable}. Notably, efforts to explain \ac{RL}-driven network management techniques are also actively researched \cite{ayoub2024towards}. In terms of network security, interpreting models for fault localization and identification have recently provided valuable insights into their operation \cite{karandin2022if,ayoub2022explainable,barnard2022robust}. However, to the best of our knowledge, no comprehensive work has yet explained physical layer attack detectors, specifically aggregated ones designed to alert against evolving threats. Therefore, this work fills that gap by providing a broad analysis of decision paths of various diverse models. 

Another important consideration is the security of the \ac{ML} models themselves. Contemporary research highlights the need to protect them against adversarial attacks, which pose a~serious risk to production models used in industry \cite{kumar2020adversarial}. In particular, attacking models designed for threat detection can have devastating consequences \cite{marek2024securing,krzyszton2024evasion}. In this context, we extend our contribution by providing, for the first time, a resilience analysis of physical layer attack detectors and examining how their \ac{XAI}-based optimization affects them.

\section{Experimental Environment and the Dataset}
\label{sec:experimental_environment}
This Section describes the experimental environment in which the data was collected. Later, the setup of the following \ac{ML} experiments is provided. 

\subsection{Data Collection}
\begin{figure}[h]
    \centering
    \includegraphics[width=\linewidth]{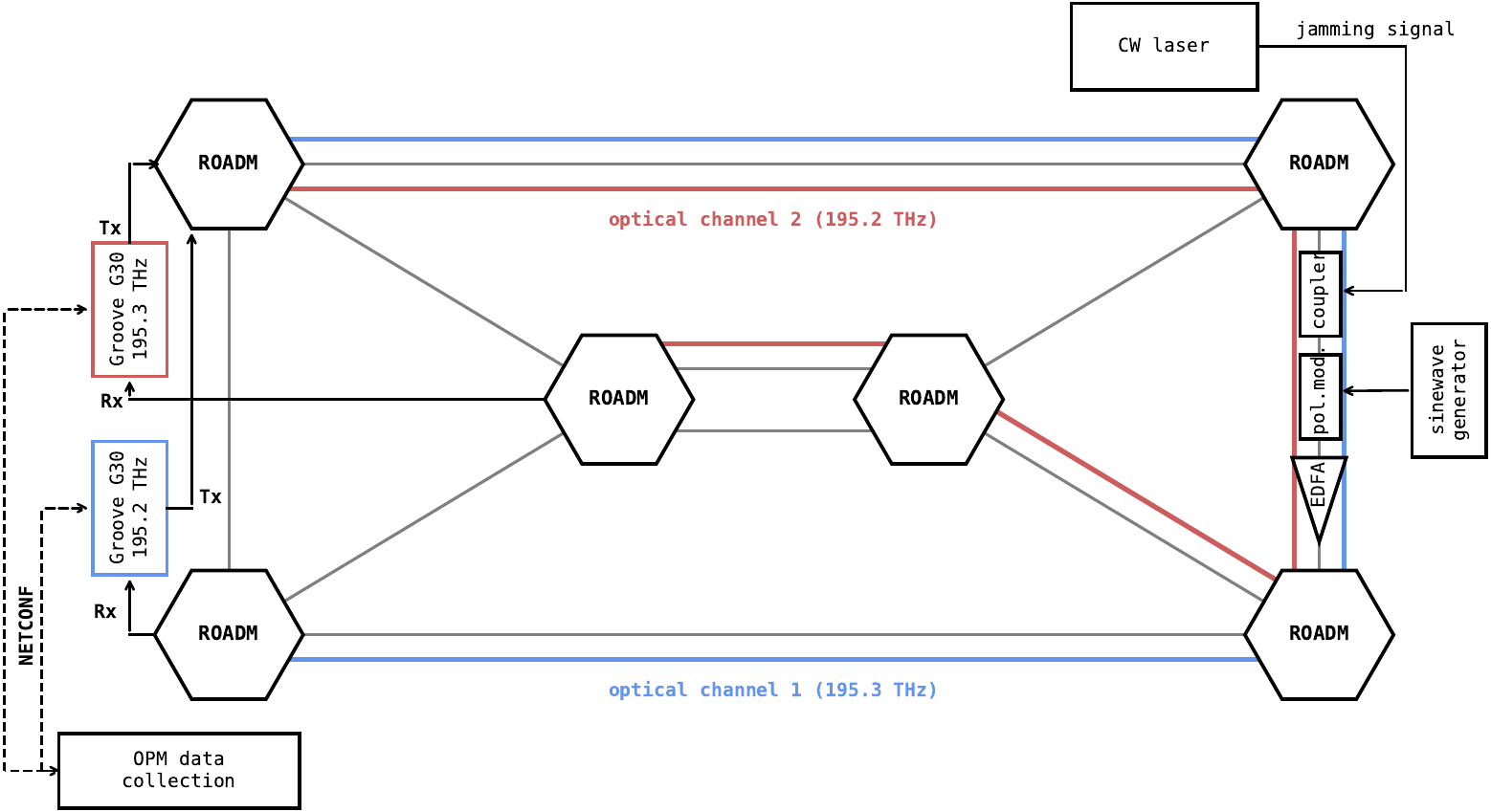}
    \caption{Scheme of the data collection experimental setup.}
    \label{fig:experimental_setup}
\end{figure}

The dataset used in this work comprises \ac{OPM} samples collected during various attacks conducted in a laboratory environment. 
The physical-layer security data used in this study was obtained through a series of experiments where an optical network testbed was subject to a set of attacks aimed at service disruption. The testbed comprises six nodes equipped with coherent transceivers and Reconfigurable Optical Add-Drop Multiplexers (ROADMs), one node with an Erbium-Doped Fiber Amplifier (EDFA) and 10 optical fiber links. The two monitored channels consist of two optical 200 Gbit/s polarization-mul\-ti\-ple\-xed 16-Quadrature Amplitude Modulation (16QAM) signals centered at 195.2 and 195.3 THz. An overview of the data collection experimental setup is illustrated in Figure \ref{fig:experimental_setup}. A detailed description of the testbed and performed experiments is provided in previous work \cite{furdek2020machine}. 

We consider three types of attacks: in-band jamming (INB), out-of-band jamming (OOB), and polarization scrambling (POL). 
In in-band jamming, a low-power jamming signal whose frequency falls within the bandwidth of the signal under test is inserted into the fiber through a passive coupler, adding unfilterable noise. We consider two power levels of the jamming signal: 10 and 7 dB below the channel under test, for a lighter (INBLGT) and a stronger (INBSTR) attack intensity, respectively.

In out-of-band jamming, the frequency of the intrusion signal is separate from the bandwidth of the channel under test and its power is higher, which causes a reduction in the amount of amplifier gain that is assigned to the legitimate channel. Again, we consider two power levels of the jamming signal: 3 and 8.7 dB above the channel under test, for a~lighter (OOBLGT) and a stronger (OOBSTR) attack intensity, respectively. 

In polarization modulation attack, a polarization modulator is clipped onto the fiber, comprising a fiber squeezer that produces stress-induced birefringence and causes polarization modulation. The induced changes are faster than the coherent receiver's polarization recovery algorithm, leading to burst errors without affecting the power or the frequency of the carried signal. The modulator is driven by a sine wave signal with 136 kHz frequency and 0.4 and 1.6 V peak-to-peak amplitude, modeling a lighter (POLLGT) and a stronger (POLSTR) attack intensity, respectively.  

\begin{table}[h]
    \centering
    \caption{\acf{OPM} Parameters of Each Data Sample}
    \begin{tabular}{p{5cm} p{3cm}}
        \toprule
        \bfseries Description & \bfseries  Acronym \\
        \midrule
        Chromatic Dispersion & CD \\
        Differential Group Delay & DGD\\
        Optical Signal to Noise Ratio & OSNR\\
        Polarization Dependent & Loss PDL\\
        Q factor & Q-factor\\
        Block Errors before FEC & BE-FEC\\
        Bit Error Rate before FEC & BER-FEC\\
        Uncorrected Block & UBE-FEC \\
        Bit Error Rate after FEC & BER-POST-FEC\\
        Optical Power Received & OPR\\
        Optical Power Transmitted & OPT\\
        Optical Frequency Transmitted & OFT\\
        Optical Frequency Received & OFR\\
        Loss Of Signal & LOS \\
        \bottomrule
        \multicolumn{2}{p{8.5cm}}{\footnotesize The maximum, minimum and average values per 1-min observation interval are reported for all parameters except BE-FEC, UBE-FEC, and LOS.
        }
    \end{tabular}
    \label{tab:opm-parameters}
\end{table}

Each attack type, as well as normal traffic, is represented by 1440 one-day traces, described using 31 distinct \ac{OPM} parameters summarized in Table \ref{tab:opm-parameters}. It is important to note that the parameters used in the dataset are those already monitored and recorded by network operators. This ensures that the detectors developed and analyzed in this work can be readily deployed in real-world environments without requiring additional instrumentation or data collection. 

\subsection{ML Experiments Setup}
To simulate a realistic scenario, the dataset was constructed with an imbalanced class distribution: for every 100 \emph{normal} samples, there are 15 \emph{attack} samples. This configuration is used across all experimental setups, including scenarios involving the detection of a single attack type and those with mixed attack types in one dataset. To ensure statistical significance, each experiment was repeated 100 times, and the average results are reported in the following sections. 

To detect the attacks, we evaluate three distinct classifiers to explore how different categories of \ac{ML} algorithms respond to diverse physical-layer intrusions. These include the neural network-based \ac{MLP}, the tree-based \ac{XGB}, and the kernel-based \ac{SVM}. We use the \texttt{scikit-learn} \cite{pedregosa2011scikit} implementations of \ac{MLP} (configured with a single hidden layer of 100 neurons, \emph{ReLU} activation, and the \emph{adam} optimizer) and \ac{SVM} (with an \emph{rbf} kernel and $C=1.0$). For \ac{XGB}, we employ the authors’ implementation \cite{chen2016xgboost}.

In previous work \cite{knapinska2025experimental}, we demonstrated that these classifiers are well-suited for detecting the considered attacks, consistently achieving balanced accuracy above 90\%. In particular, our experiments showed that specialized detectors trained for a~specific attack type were able to detect such attacks with near-perfect consistency. Meanwhile, models trained on aggregated data from all attack types also performed well, achieving high accuracy even on previously unseen attacks. However, the inner workings of these models remained unexplored. 

In the following sections, we analyze their behavior in greater detail to gain deeper insights. The resulting knowledge will help identify opportunities for optimization, particularly in scenarios where certain parameters may be obsolete or irrelevant. Specifically, we investigate how model behavior varies depending on the attack types used during training. 

\section{Understanding Attack Detectors}
\label{sec:xai_analysis}
To gain insight into the internal operation of the attack detectors we use \ac{SHAP} – a game-theoretic framework that estimates the contribution of individual features to the overall output of a \ac{ML} model. 

\begin{figure}[h]
    \centering
    \includegraphics[width=\linewidth]{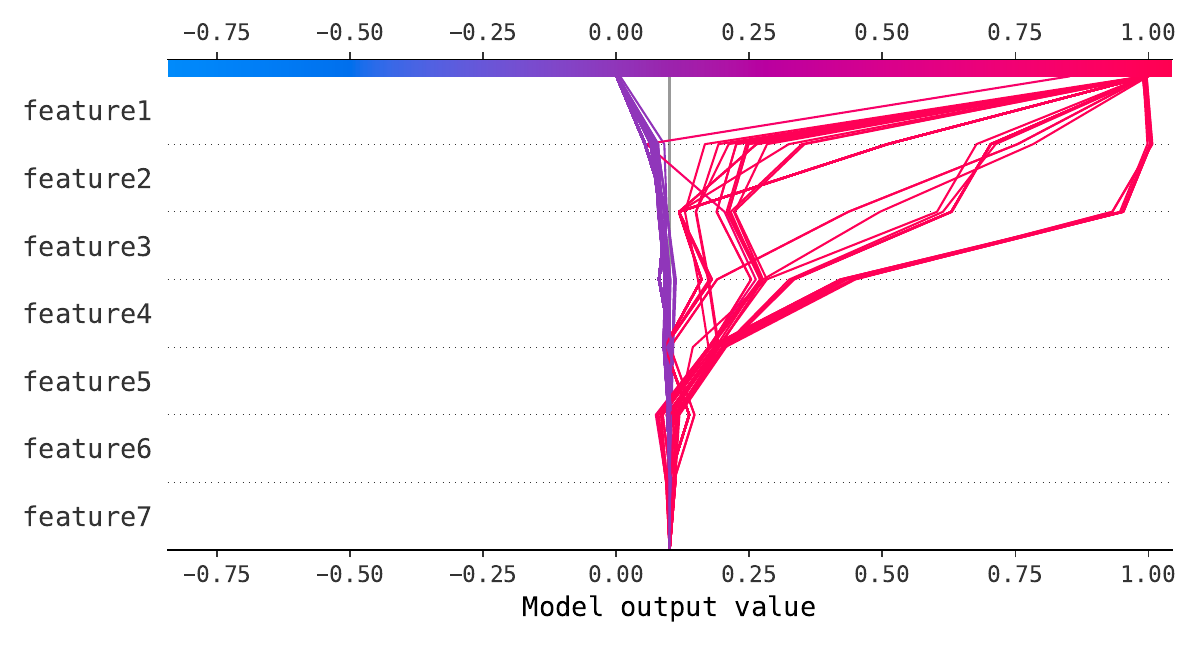}
    \caption{Example \ac{SHAP} decision plot for the attack detection task. }
    \label{fig:example_decision_plot}
\end{figure}

We analyze \emph{decision plots}, which illustrate how the model arrives at its final decision by considering various features and their respective contributions to the overall prediction. Consider an example in Figure \ref{fig:example_decision_plot}. The features on the left-hand side of each plot are ordered by their influence on the model’s final decision – from most to least influential, top to bottom. The x-axis represents the predicted probability that a~given sample from the dataset corresponds to an attack. The gray vertical line indicates the base prediction (i.e., the model’s output if all features had random values). In all cases analyzed in this study, this base value is approximately 0.1, suggesting that the model considers the likelihood of an attack to be very low by default. Consequently, the model returns a negative (no attack) decision when uncertain. 

The colored lines show the model’s step-by-step reasoning for individual samples; each line corresponds to one sample from the dataset. By following a line from the bottom of the plot upward, we can trace the decision-making process. The decision starts at the base value and is then adjusted feature by feature. As we move up, each feature either increases or decreases the predicted probability that the sample describes an attack, shifting the decision in a positive (attack) or negative (no attack) direction. 

Note that the \ac{ML} models operate on probabilities that combine to produce a final decision. As a result, in the lower parts of the plot (before all features have been analyzed), negative probabilities might appear, as the cumulative probability at a~given point can drop below 0. 

Another important consideration is that the plots show samples after classification, with red lines representing the attack class, and purple lines representing the non-attack class. As a result, a feature, even if it appears high in the ranking, may influence the prediction only for samples from one of the classes. 

\begin{figure*}[!ht]
    \centering
    \includegraphics[width=.3\linewidth]{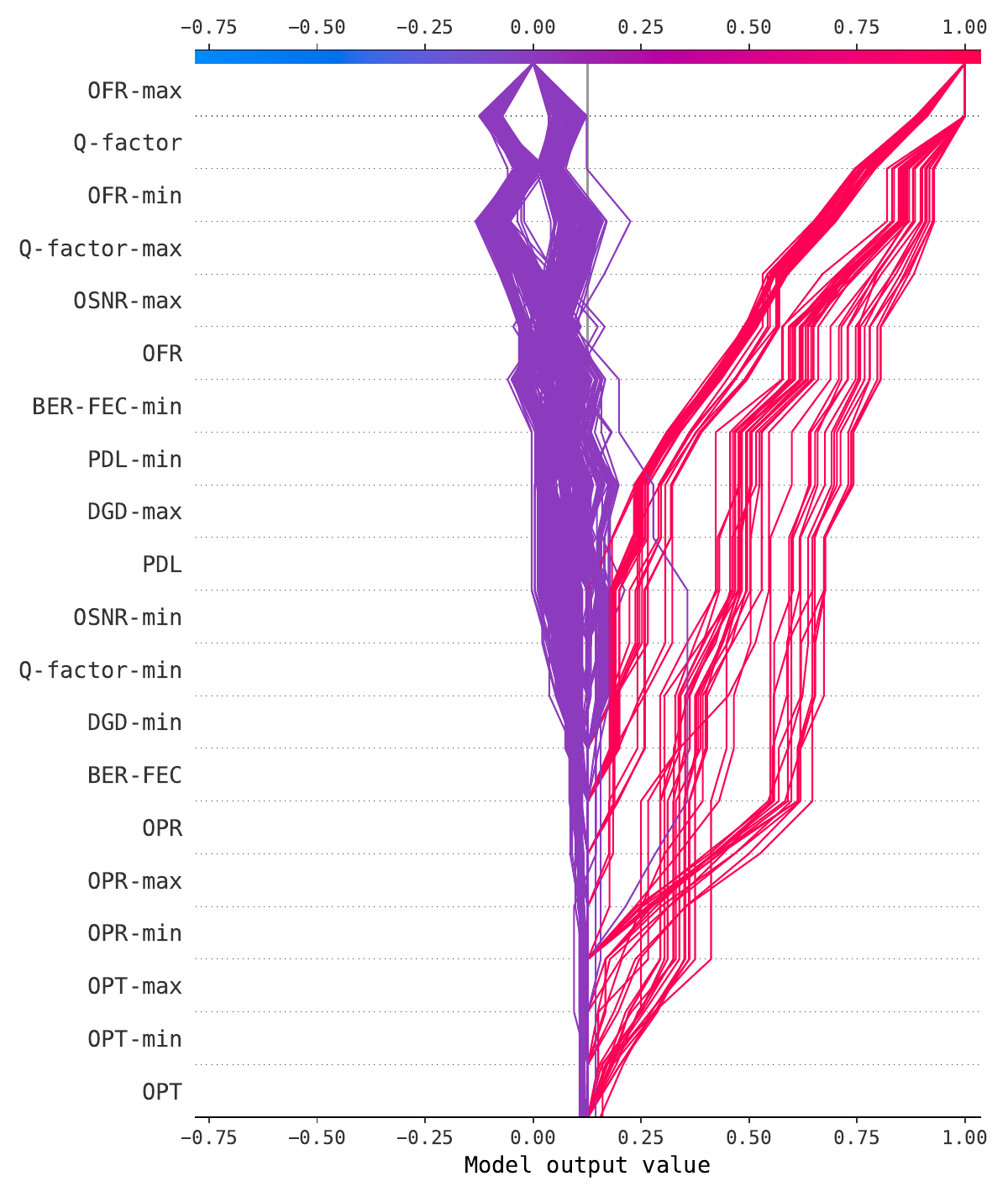}
    \includegraphics[width=.3\linewidth]{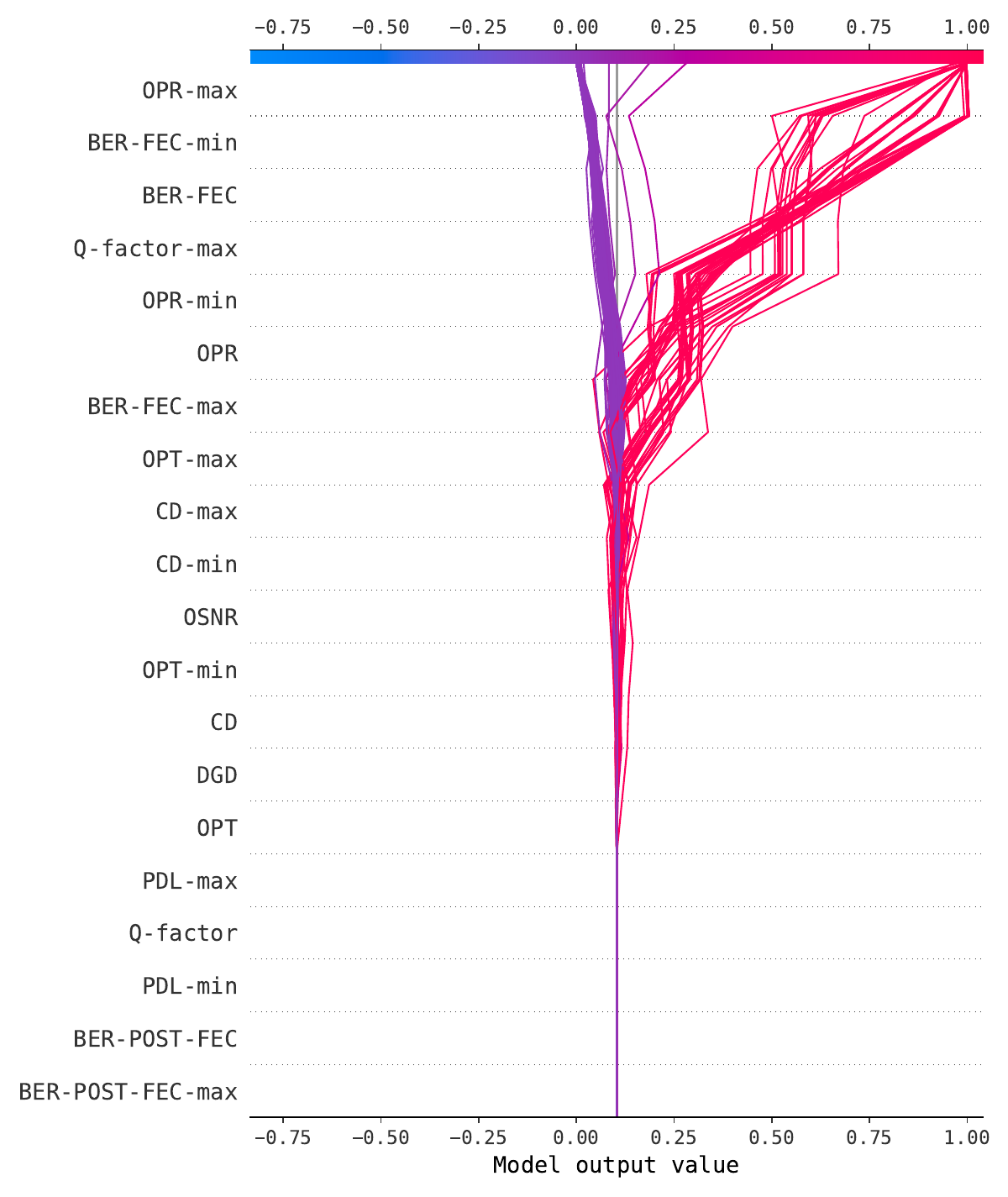}
    \includegraphics[width=.3\linewidth]{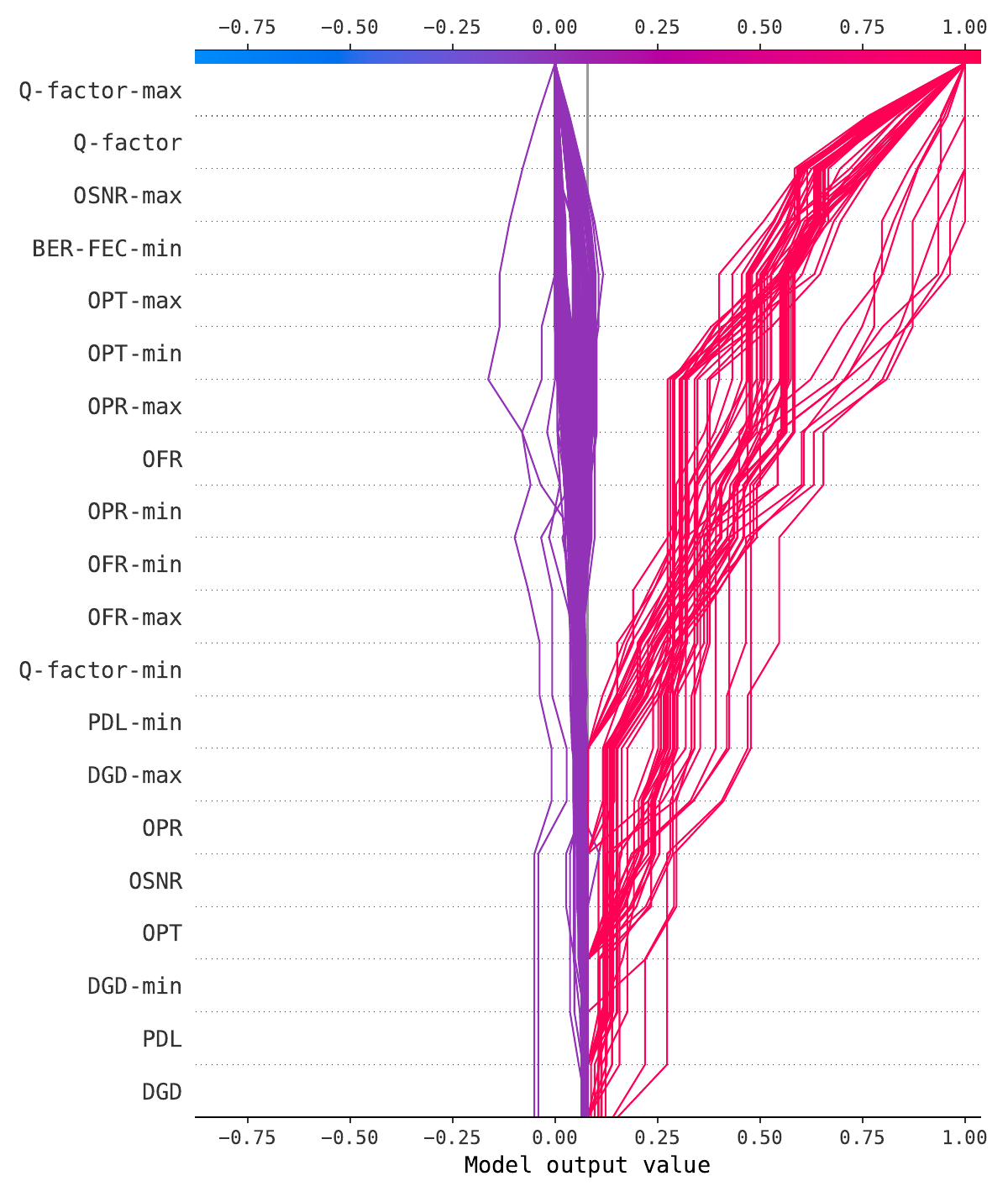}
    \includegraphics[width=.3\linewidth]{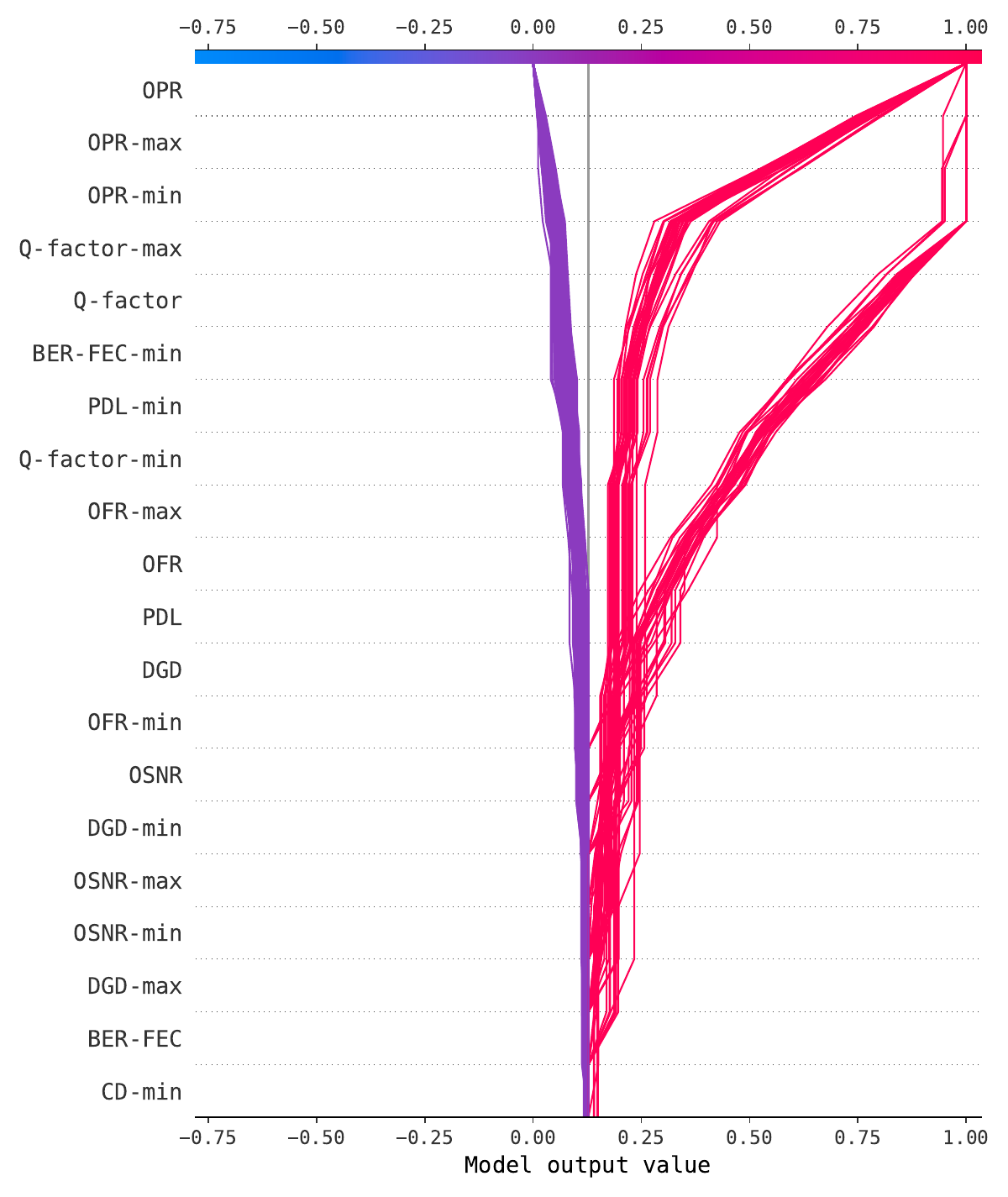}
    \includegraphics[width=.3\linewidth]{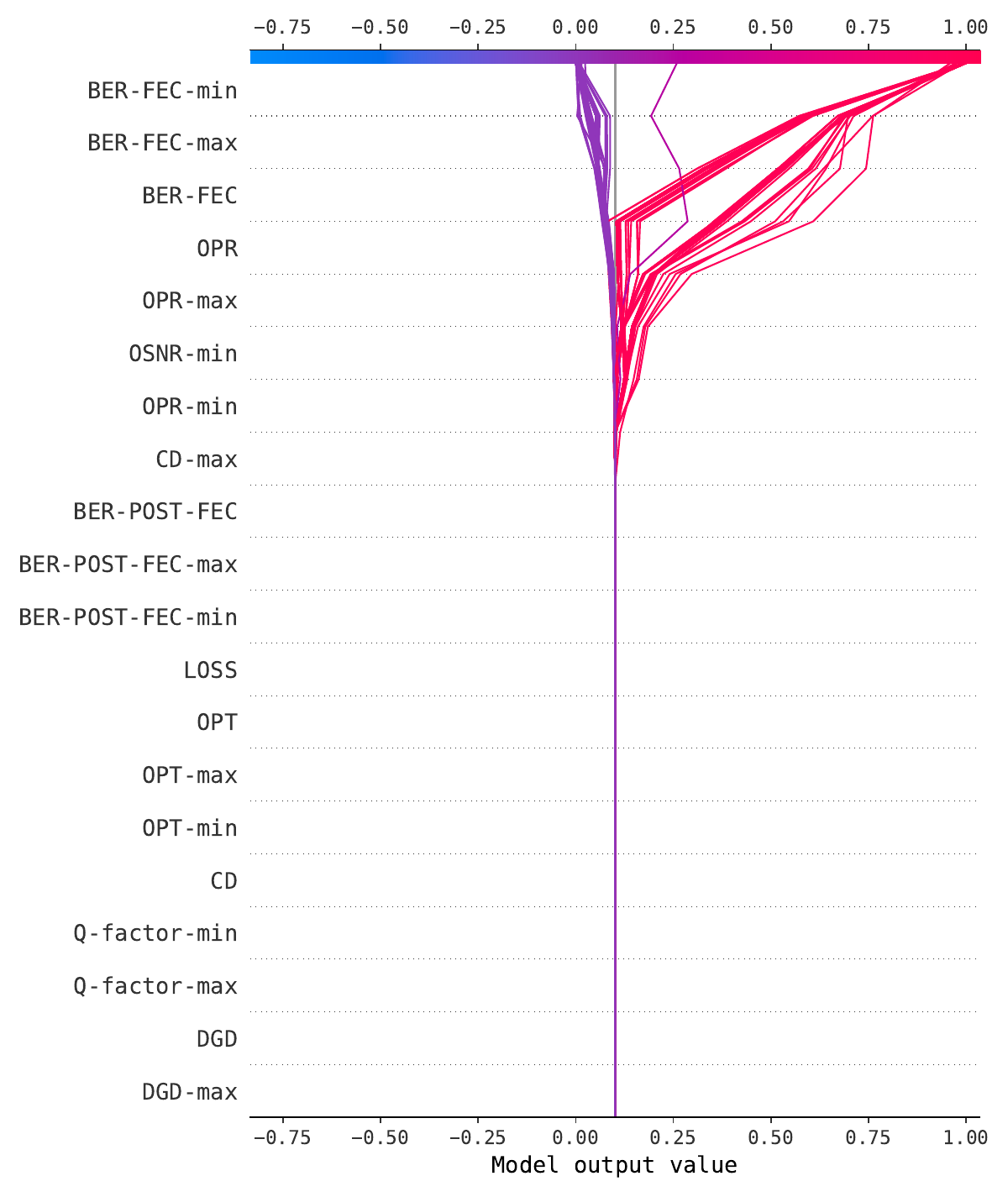}
    \includegraphics[width=.3\linewidth]{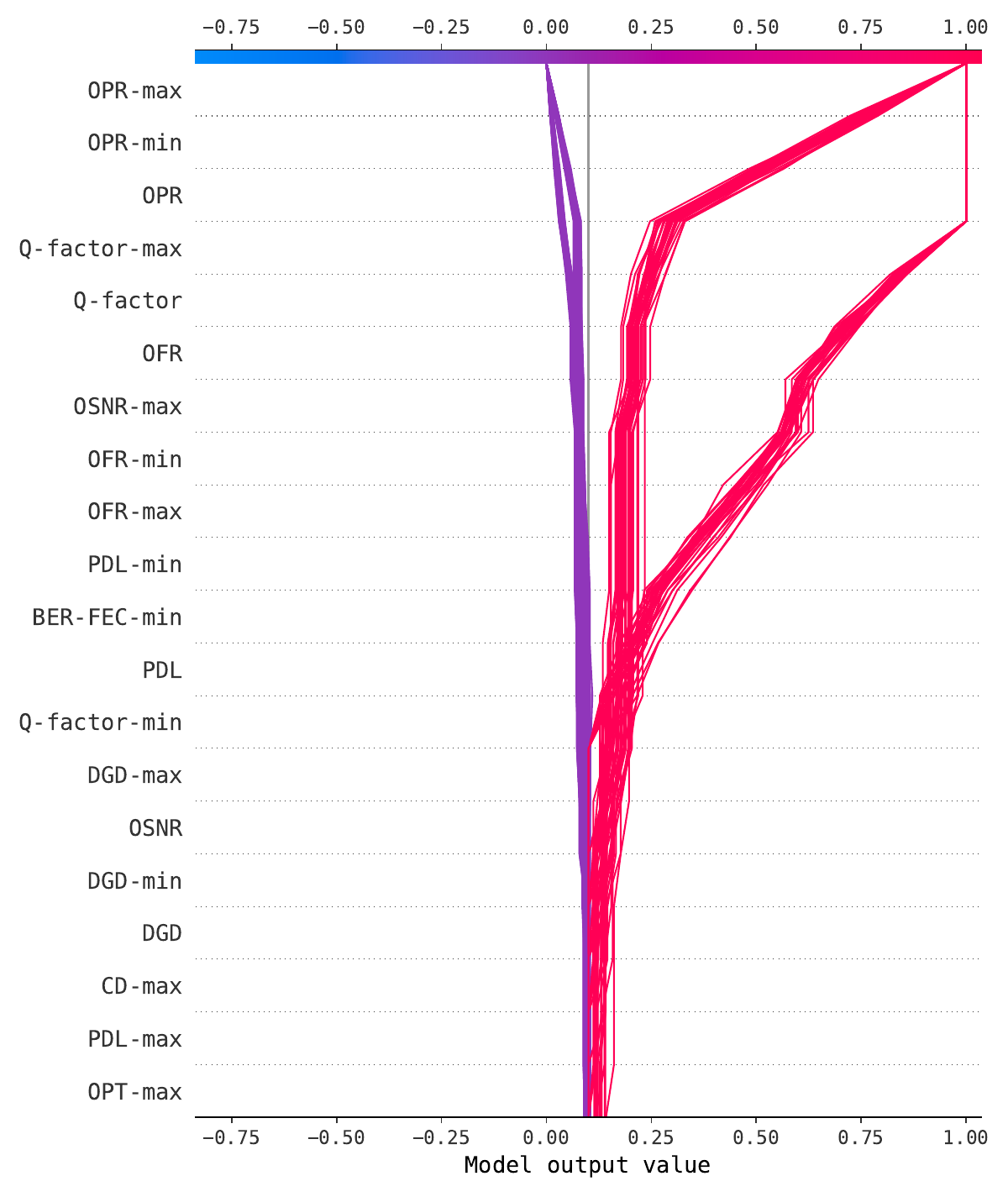}
    \caption{SHAP summary plots for models detecting INBLGT (first row) and INBSTR (second row). MLP (right), XGB (middle), and SVM (left) classifiers. }
    \label{fig:shap_inb}
\end{figure*}

In the following part, we analyze the operation of dedicated detectors in Section \ref{ssec:dedicated_detectors}, and the aggregated detectors in Section \ref{ssec:aggregated_detectors}. 

\subsection{Dedicated Detectors}
\label{ssec:dedicated_detectors}
As the number of analyzed cases is quite large (six attack types, three types of detectors per each of them), let us consider a few illustrative examples. Figure \ref{fig:shap_inb} presents the \ac{SHAP} decision plots of three types of classifiers (\ac{MLP}, \ac{SVM}, \ac{XGB}) detecting the light and strong in-band jamming attack (INBLGT and INBSTR, respectively). 

Analyzing the INBLGT attack detection, we observe that each classifier recognizes it differently, with feature rankings varying across models. However, the Q-factor consistently appears among the top features in all three classifiers, suggesting it plays a key role in detecting this particular attack. 

Notably, both the \ac{MLP} and \ac{SVM} rely on multiple pa\-ra\-me\-ters to form their decisions, integrating information from a wide range of features and considering a broader set of network parameters in their decision-making process. In contrast, the \ac{XGB} model appears to depend primarily on just six features. This difference likely stems from the underlying algorithmic design: \ac{XGB} is based on decision trees, which use features as splitting criteria and naturally highlight those with the highest information gain. 

In the case of the INBSTR attack, all three classifiers exhibit two distinct decision paths leading to its detection. This is evi\-dent from the red lines (representing attack decisions), which consistently follow two separate trajectories. For both the \ac{MLP} and \ac{SVM}, the decision "bending point" appears at either the Q-factor or OPR – after analyzing these parameters, the predicted probability of an attack increases rapidly. Depending on the path taken, this probability jumps to either around 0.25 or close to 1. In the case of \ac{XGB}, it is typically associated with the maximum or minimum recorded value of BER-FEC. 

In contrast, there are no strong indicators for the 'no attack' decision in any of the models; they appear to default to a~baseline classification when all parameter values remain within expected ranges. 

Once again, the feature rankings differ across the evaluated \ac{ML} algorithms. However, in this case, one parameter – OPR – appears in the top-4 for all models displayed on the plots, highlighting its consistent relevance across classifiers. 

Interestingly, although both analyzed cases involve the same type of attack (in-band jamming), they differ in intensity – lighter in the first case and stronger in the second – which leads to notably different detection patterns. This suggests that classifiers trained only on one intensity level develop quite different reasoning pathways for detection, even when the underlying attack type remains the same. 

The analysis and plots for the remaining attack types – omitted here for brevity and space constraints – reveal additional, noteworthy insights. A consistent trend emerges: (i) each attack type has its own set of features that are most critical for its detection; (ii) the intensity of an attack encountered during training influences how a classifier detects it; and (iii) each classifier employs a different strategy to detect the same attack. Recall that in each case, the models make a~binary (attack vs. no attack) decision based on the dataset used. 

The remaining plots, along with additional supplementary material, are available in the project's public GitHub repository\footnote{\url{https://github.com/w4k2/explainable_attack_detectors}}.


\begin{figure*}[!ht]
    \centering
    \includegraphics[width=.3\linewidth]{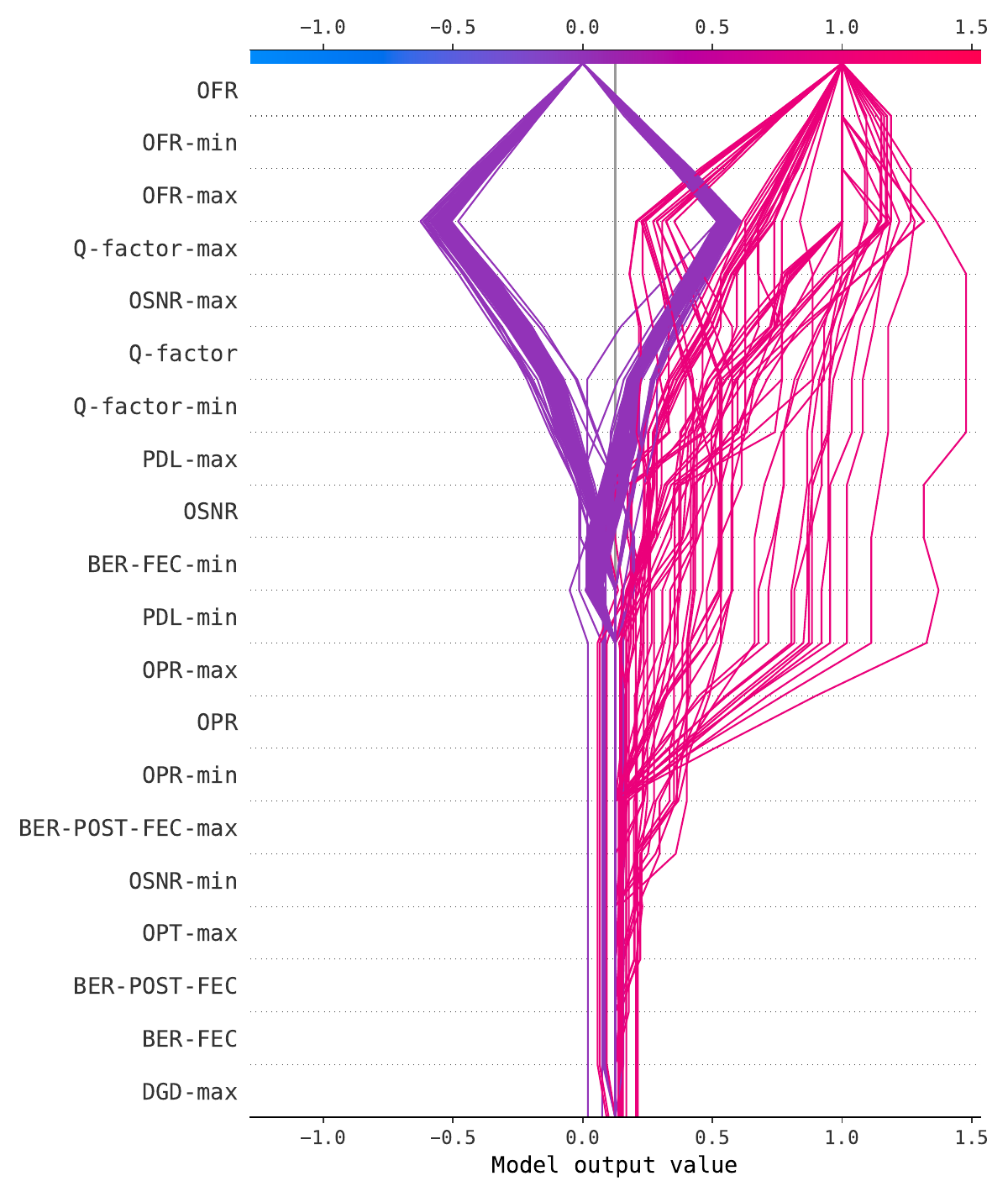}
    \includegraphics[width=.3\linewidth]{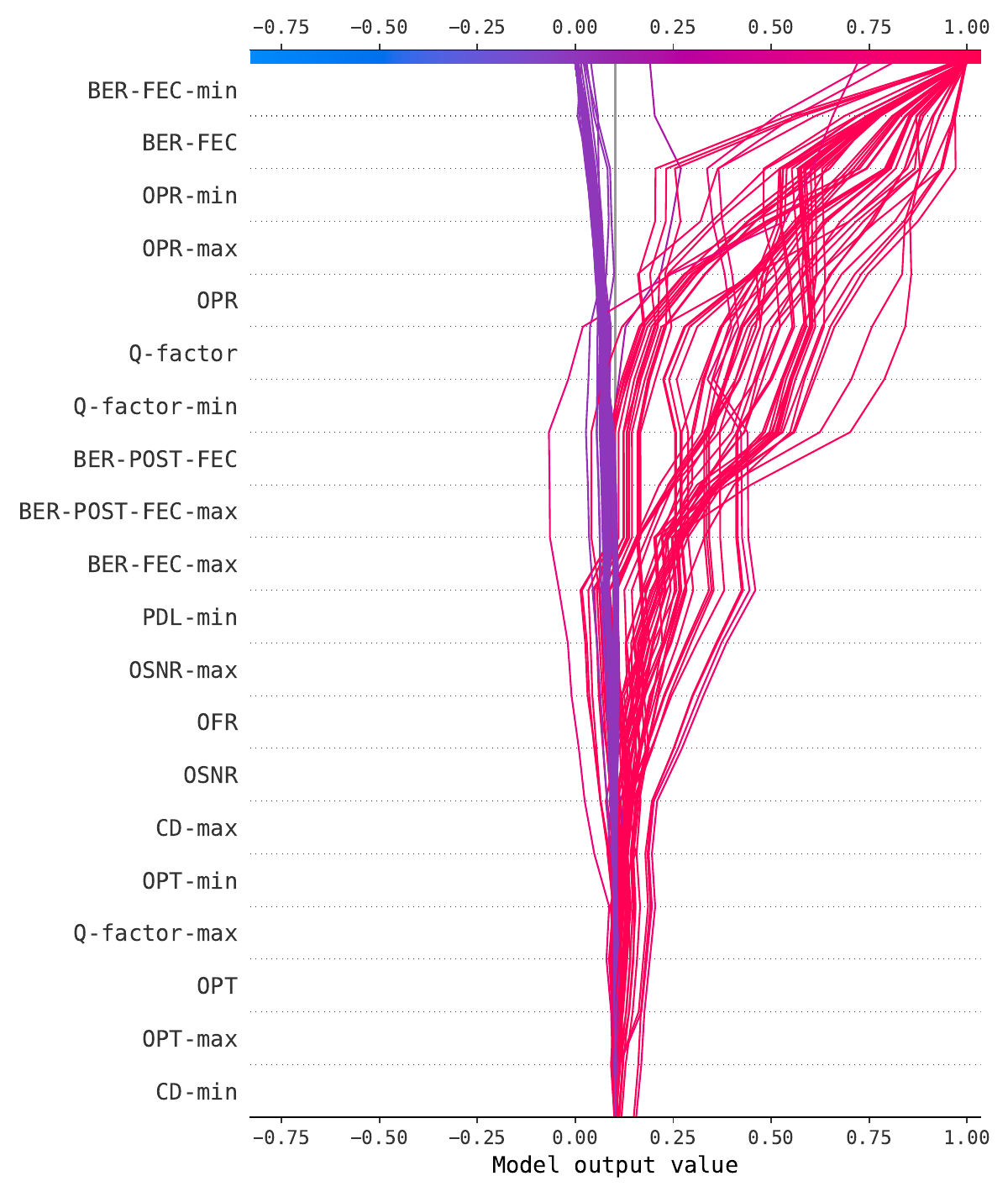}
    \includegraphics[width=.3\linewidth]{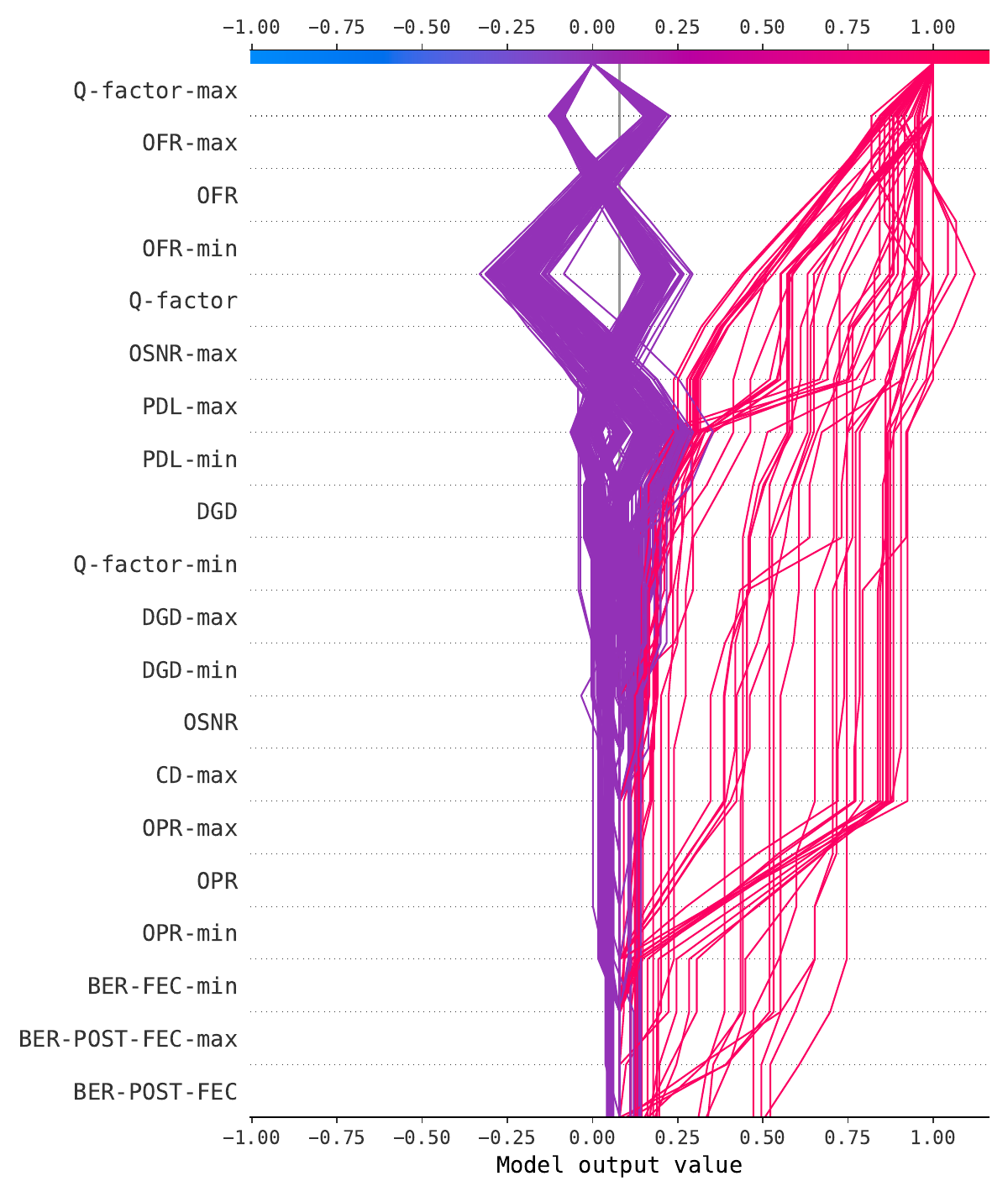}
    \caption{SHAP summary decision plots for models trained on aggregated data of all attacks. MLP (left), XGB (middle), and SVM (right) classifiers. }
    \label{fig:shap_joint_attacks}
\end{figure*}

\subsection{Aggregated Detectors}
\label{ssec:aggregated_detectors}
The above analysis highlights how the internal reasoning of each model varies depending on the specific attack scenario encountered during training. However, previous research \cite{knapinska2025experimental} demonstrated that it is possible to effectively detect attacks using an aggregated model trained on a joint dataset that combines information from multiple attack types. In the following section, we analyze how such detectors operate. 

Figure~\ref{fig:shap_joint_attacks} presents the \ac{SHAP} summary plots for models detecting anonymized attacks from aggregated data. Note that these models are trained on binary datasets (attack vs. no-attack), without distinguishing between individual attack types. 

Analyzing the decision plots for \ac{MLP} and \ac{SVM}, we observe that – unlike the dedicated detectors discussed earlier – there are two distinct decision paths leading to a 'no attack' classification. This highlights the increased complexity of training aggregated models and the greater number of features that must be considered jointly for the classifiers to make confident decisions. 

For example, note how both the Q-factor-max and OFR values are critical for the \ac{MLP} to correctly identify the absence of an attack in the current set of \ac{OPM} parameters. Furthermore, the \ac{SVM} relies on at least three parameters – including OFR-max, Q-factor, and PDL-min – which highly influence the final 'no attack' decision. This is evident in the plots, where these parameters correspond to the most prominent bends along the purple decision paths. 

Finally, the tree-based \ac{XGB} classifier also exhibits a sig\-ni\-fi\-cant\-ly more complex decision-making process when trained on aggregated datasets compared to dedicated ones. In this case, multiple features act in combination to shift its decision toward classifying a sample as a physical-layer attack. 

Although each classifier constructs its decision in a more complex manner when trained on the aggregated dataset, not all features are utilized. This suggests that, while more information is indeed required for accurate detection, some features remain unused or possibly redundant. This indicates potential for the reduction of the detectors in terms of size and, thus, possible improvements in their speed. 

\section{XAI-Based Detector Optimization}
\label{sec:detector_optimization}
The insights gained from the decision plots can be leveraged to optimize the detectors through \ac{FS}. To this end, we performed \ac{FS} based on the most influential features identified for each individual attack type, as well as for the aggregated detectors. By analyzing each decision plot, we selected a single feature set covering all attack types, focusing on features that most strongly influence the decision boundary – typically 1–3 features per attack type. Since feature importance rankings vary significantly across classifiers, we repeated this procedure for each classifier to obtain reduced, classifier-specific feature sets. The selected features for each case are listed in Table \ref{tab:selected_features}, sorted alphabetically. 

\begin{table}[!h]
    \centering
    \caption{Selected Features for Various Classifiers after \ac{XAI} Analysis}
    \begin{tabular}{l p{6.75cm}}
        \toprule
        \bfseries Classifier & \bfseries Selected Features \\
        \midrule
        MLP & BER-POST-FEC-max, OPR, OPR-min, OFR, OFR-max, OFR-min, OSNR-max, OPT-max, Q-factor, Q-factor-max, Q-factor-min \\
        XGB & BER-FEC, BER-FEC-max, BER-FEC-min, BER-POST-FEC-max, OPR, OPR-max, OPR-min, OFR, OPT, Q-factor-max \\
        SVM & BER-POST-FEC, BER-POST-FEC-max, OFR, OFR-max, OFR-min, OPR-max, OPT-max, OSNR-max, PDL-max, Q-factor, Q-factor-max, Q-factor-min \\
        \bottomrule
    \end{tabular}
    \label{tab:selected_features}
\end{table}

\begin{table*}[!hb]
\centering
\caption{Attack Detection Quality for Dedicated and Aggregated Models Before and After \ac{FS}}
\begin{tabular}{llcccccc}
\toprule
\textbf{Attack Type} & \textbf{Classifier} &
\multicolumn{2}{c}{\textbf{BAC}} &
\multicolumn{2}{c}{\textbf{F1}} &
\multicolumn{2}{c}{\textbf{G-Mean}} \\
\cmidrule(lr){3-4} \cmidrule(lr){5-6} \cmidrule(lr){7-8}
& & Orig. & \ac{FS} & Orig. & \ac{FS} & Orig. & \ac{FS} \\
\midrule
INBLGT  & MLP & 0.9914 & 0.9779 & 0.9867 & 0.9720 & 0.9913 & 0.9777 \\
        & XGB & 0.9900 & 0.9839 & 0.9858 & 0.9781 & 0.9900 & 0.9838 \\
        & SVM & 0.9368 & 0.9201 & 0.9312 & 0.9083 & 0.9347 & 0.9168 \\
INBSTR  & MLP & 0.9998 & 0.9998 & 0.9998 & 0.9998 & 0.9998 & 0.9998 \\
        & XGB & 0.9985 & 0.9982 & 0.9963 & 0.9966 & 0.9985 & 0.9982 \\
        & SVM & 0.9998 & 0.9998 & 0.9998 & 0.9998 & 0.9998 & 0.9998 \\
OOBLGT  & MLP & 0.9994 & 0.9995 & 0.9994 & 0.9995 & 0.9994 & 0.9995 \\
        & XGB & 0.9992 & 0.9996 & 0.9976 & 0.9985 & 0.9992 & 0.9996 \\
        & SVM & 0.9638 & 0.9794 & 0.9625 & 0.9790 & 0.9632 & 0.9792 \\
OOBSTR  & MLP & 0.9972 & 1.0000 & 0.9971 & 1.0000 & 0.9972 & 1.0000 \\
        & XGB & 0.9983 & 0.9977 & 0.9981 & 0.9976 & 0.9983 & 0.9977 \\
        & SVM & 0.9924 & 1.0000 & 0.9923 & 1.0000 & 0.9923 & 1.0000 \\
POLLGT  & MLP & 0.9968 & 0.9982 & 0.9967 & 0.9982 & 0.9968 & 0.9982 \\
        & XGB & 0.9970 & 0.9966 & 0.9966 & 0.9964 & 0.9970 & 0.9966 \\
        & SVM & 0.9946 & 1.0000 & 0.9946 & 1.0000 & 0.9946 & 1.0000 \\
POLSTR  & MLP & 0.9981 & 0.9981 & 0.9981 & 0.9981 & 0.9981 & 0.9981 \\
        & XGB & 0.9959 & 0.9959 & 0.9959 & 0.9959 & 0.9959 & 0.9959 \\
        & SVM & 1.0000 & 1.0000 & 1.0000 & 1.0000 & 1.0000 & 1.0000 \\
\rowcolor{red!10}
aggregated   & MLP & 0.9849 & 0.9851 & 0.9847 & 0.9849 & 0.9848 & 0.9850 \\
\rowcolor{red!10}
        & XGB & 0.9937 & 0.9939 & 0.9919 & 0.9914 & 0.9937 & 0.9939 \\
\rowcolor{red!10}
        & SVM & 0.9136 & 0.9293 & 0.9054 & 0.9239 & 0.9095 & 0.9266 \\
\bottomrule
\end{tabular}
\label{tab:comparison_results_fs}
\end{table*}

Table~\ref{tab:comparison_results_fs} presents a comparison between the performance of the original models trained on the full feature set and those trained after applying \ac{FS}. Following the approach from \cite{knapinska2025experimental}, we consider three evaluation metrics tailored for imbalanced classification problems: \ac{BAC}, \ac{F1}, and \ac{G-Mean}. 

\begin{figure*}[h]
    \centering
    \includegraphics[width=\linewidth]{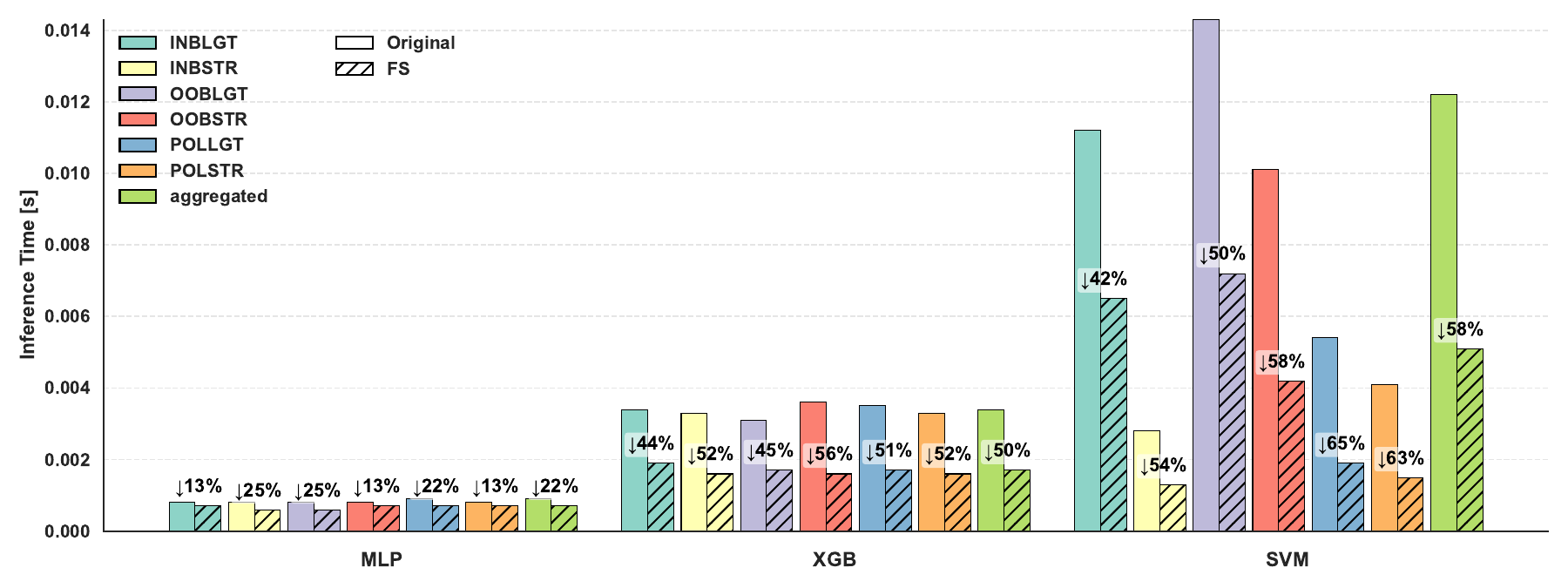}
    \caption{Inference time of the classifiers before and after \ac{FS} for the considered physical layer attack types.}
    \label{fig:inference_time_fs}
\end{figure*}

The first notable trend is the near-perfect effectiveness of the dedicated detectors trained on the full feature set. However, the drop in performance when using aggregated data from all attack types is minimal, especially for \ac{XGB} and \ac{MLP} (see the highlighted rows at the bottom of the table). Remarkably, after applying \ac{FS} and retaining only the most influential features, performance metrics improved in nearly all cases when aggregating data from all attack types into a single dataset. The best-performing algorithm, \ac{XGB}, improved from 0.9937 to 0.9939 in \ac{BAC}, while the weakest, \ac{SVM}, improved from 0.9136 to 0.9293. This demonstrates that reducing the number of parameters used for training not only preserves, but often enhances, model performance. These improvements are evident in both the dedicated and aggregated models. 

In terms of processing time – both training and inference – another important parameter influenced by \ac{FS}, the benefits are significant. Training time was reduced by up to 36\% for \ac{MLP}, 95\% for \ac{XGB}, and 55\% for \ac{SVM}, depending on the attack type. Detailed results are available in the public GitHub repository. However, a more critical parameter in practical applications is inference time, which was also significantly reduced as a result of \ac{FS}. Inference time measures how quickly the model makes a decision, i.e., how fast an alarm can be raised after processing a sample. Figure~\ref{fig:inference_time_fs} presents the improvements for the evaluated classifiers and attack types. An analysis of the plot reveals that even the fastest-inferencing model, \ac{MLP}, achieved a speed-up of up to 25\%. \ac{XGB} became twice as fast, while \ac{SVM} saw a reduction in inference time of up to 64\%.

\section{Resilience of the Detectors}
\label{sec:detector_resilience}
The \ac{FS} procedure allowed us to significantly reduce the size of the attack detectors, making them noticeably faster without sacrificing quality. However, this reduction might also make them more vulnerable to instability and less resilient to poisoning attacks \cite{kumar2020adversarial}. Specifically, malicious actors could deliberately tamper with input measurements, preventing the detectors from raising alarms when anomalies are present. In such cases, the models would have less backup information from the remaining parameters, potentially compromising their accuracy. 

Therefore, to test the resilience of the optimized detectors, we performed a parameter noising experiment. To this end, for each attack, we replaced the values of the most influential feature (top one in the ranking on the plot) with noise. We repeated this experiment for each of the considered classifiers. The complete list of features noised in each case is provided in Table \ref{tab:noised_features}. For feature groups representing statistics of a~single parameter (e.g., Q-factor, Q-factor-max, Q-factor-min), the entire group was noised to ensure consistency in the representation of that parameter. To keep the table concise, only the main feature in each group is listed. The detailed results of these experiments are available in the public GitHub repository; below, we discuss a summary of the key findings. 

\begin{table}[!h]
    \centering
    \caption{Noised Features for Various Attack Types and Classifiers}
    \begin{tabular}{lccc}
        \toprule
        \bfseries Attack Type & \bfseries MLP & \bfseries XGB & \bfseries SVM \\
        \midrule
        INBLGT & OFR & OPR & Q-factor \\
        INBSTR & OPR & BER-FEC & OPR \\
        OOBLGT & OFR & OPR & Q-factor \\
        OOBSTR & OFR & Q-factor & Q-factor \\
        POLLGT & Q-factor & BER-POST-FEC & Q-factor \\
        POLSTR & BER-POST-FEC & BER-POST-FEC & Q-factor \\
        aggregated & OFR & BER-FEC & Q-factor \\
        \bottomrule
    \end{tabular}
    \label{tab:noised_features}
\end{table}

The first clear trend observed is that introducing noise into the most important \ac{OPM} parameters significantly reduces the models’ ability to detect attacks, even when the full feature set is used during training. This suggests that, once the most influential feature is corrupted, the models are forced to rely on other, less informative parameters. As a result, the detection task becomes more difficult and less reliable. 

This effect is even more pronounced for detectors optimized through \ac{FS}. The drop in detection performance – when moving from the original to the noised parameter set – is substantially larger in these cases. We can therefore conclude that, although the reduced models are considerably faster, they are also less resilient to input perturbations such as feature noising. 

\begin{figure}
    \centering
    \includegraphics[width=\linewidth]{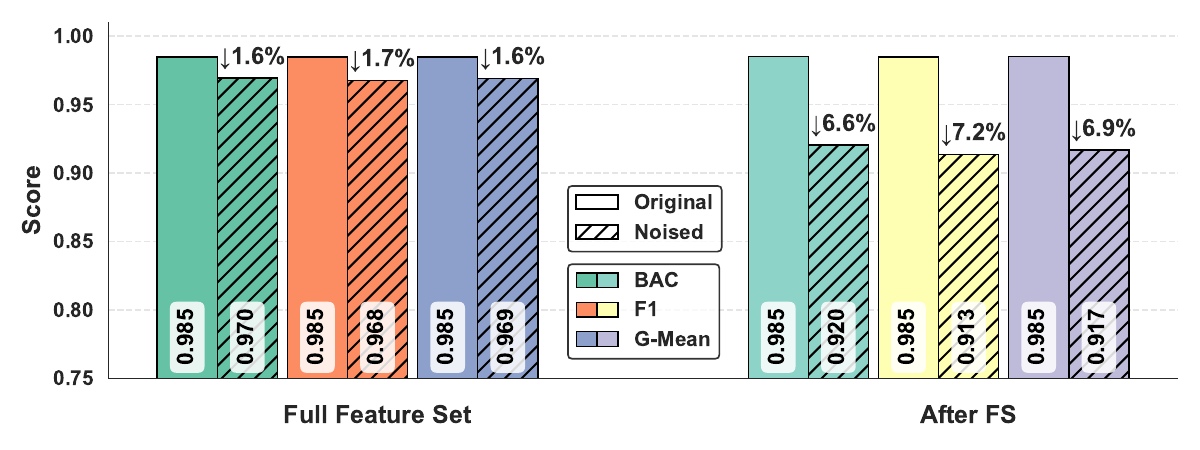}
    \caption{Attack detection quality according to various metrics for the original and noised feature values; aggregated detector; \ac{MLP} classifier. }
    \label{fig:mlp_metric_comparison_noising}
\end{figure}

An illustrative example is shown in Figure~\ref{fig:mlp_metric_comparison_noising}, which presents the detection performance of the \ac{MLP} model trained on aggregated data. The drop in quality after noising is significantly more severe when \ac{FS} has been applied. Specifically, on the left-hand side of the plot, which shows the model trained on the full feature set, the quality drop is below 2\%. In contrast, on the right-hand side of the plot, showing the model's performance after \ac{FS}, we observe a quality loss of around 7\%.

Table~\ref{tab:parameter_noising} summarizes the results of the experiment, comparing the average detection quality drop for models trained on the full feature set and those trained after \ac{FS}, across each attack type, classifier, and overall. Although the magnitude of the drop varies, in all cases, models trained on the full feature set demonstrated greater resilience to malicious parameter noising. For example, in the case of the INBLGT attack detection, we observe a 7.82\% drop in \ac{BAC} after parameter noising when using the full feature set. After applying \ac{FS}, this drop increases to 9.35\%. A much more severe example occurs for the detection of the INBSTR attack – the drop in \ac{BAC} increases from 0.84\% to 9\%. 
Among the classifiers, \ac{MLP} is the most stable when using the full feature set, while \ac{XGB} shows the greatest resilience after \ac{FS} is applied. \ac{SVM} experiences a significant loss in detection quality, particularly when optimized through \ac{FS}.

\begin{table*}
\centering
\caption{Decrease of Performance Metric [\%] with and without Parameter Noising}
\begin{tabular}{@{}llccc|ccc@{}}
\toprule
\textbf{Category} & \textbf{Type} & \multicolumn{3}{c|}{\textbf{Full Feature Set}} & \multicolumn{3}{c}{\textbf{After \ac{FS}}} \\
\cmidrule(lr){3-5} \cmidrule(l){6-8}
 & & BAC & F1 & G-Mean & BAC & F1 & G-Mean \\
\midrule
\multirow{7}{*}{Attack Type} 
& INBLGT & 7.82 & 10.21 & 8.73 & 9.35 & 12.97 & 10.85 \\
& INBSTR & 0.84 & 1.04 & 0.85 & 9.00 & 10.87 & 9.81 \\
& OOBLGT & 4.92 & 6.11 & 5.52 & 6.26 & 7.90 & 7.08 \\
& OOBSTR & 5.58 & 6.60 & 6.07 & 6.67 & 7.80 & 7.23 \\
& POLLGT & 2.73 & 2.99 & 2.85 & 3.08 & 3.35 & 3.20 \\
& POLSTR & 0.23 & 0.24 & 0.24 & 0.44 & 0.45 & 0.44 \\
& aggregated  & 3.85 & 5.07 & 4.44 & 5.48 & 6.68 & 6.08 \\
\midrule
\multirow{3}{*}{Classifier} 
& MLP     & 1.00 & 1.17 & 1.01 & 3.13 & 3.49 & 3.25 \\
& SVM     & 8.70 & 10.87 & 9.78 & 12.54 & 15.80 & 14.20 \\
& XGB     & 1.43 & 1.79 & 1.51 & 1.59 & 2.15 & 1.70 \\
\midrule
\textbf{Overall Avg.} & -- & 3.71 & 4.61 & 4.10 & 5.75 & 7.15 & 6.39 \\
\bottomrule
\end{tabular}
\label{tab:parameter_noising}
\end{table*}

These findings reveal a clear trade-off between the two experimental outcomes. On one hand, \ac{XAI}-based \ac{FS} allows for substantial optimization of the attack detectors – making them up to twice as fast without compromising detection quality under normal conditions. On the other hand, the optimized detectors proved to be less robust when exposed to adversarial input perturbations, such as parameter noising. 

Thus, our study highlights an important design consideration when constructing attack detectors based on available \ac{OPM} parameters: balancing the goals of computational efficiency and adversarial robustness, which may often be at odds. Therefore, future model designs should take this trade-off into account and adjust according to the operating conditions, aiming to achieve both high performance and resilience.

\section{Conclusions}
\label{sec:conclusions}
In this paper, we focused on the detection of physical-layer attacks in communication networks. Using a real, ex\-pe\-ri\-men\-ta\-lly collected dataset, we constructed a set of dedicated detectors for six types of attacks. Additionally, we considered an aggregated scenario, in which a single detector is trained on data from all available attack types to raise alarms. 

Leveraging \ac{XAI} tools, we performed an in-depth analysis of the internal decision-making processes of the models, uncovering several noteworthy insights. The generated decision plots revealed that the same \ac{OPM} parameters are used differently depending on the type of attack and the \ac{ML} algorithm employed. Based on these findings, we applied \ac{FS} to optimize the detectors. This process allowed us to significantly reduce model size and improve inference speed – by up to a factor of two – without compromising detection quality. 

Finally, we evaluated the resilience of the developed detectors against adversarial perturbations through a parameter noising experiment. By replacing the most influential features with noise, we measured the resulting drop in detection performance. This experiment exposed a critical trade-off: while the original detectors demonstrated greater robustness to parameter noising, the optimized models, though faster, were less resilient. These findings underscore two opposing design considerations when building attack detectors based on available \ac{OPM} parameters: substantial speed improvements are possible, but they may come at the cost of reduced robustness to adversarial manipulation. Therefore, the guidelines provided in this work support informed decision-making for designing attack detectors based on the specific operational requirements of each network. 

In the future, we plan to extend our research by assessing the impact of parameter noising on dynamically updated attack detectors in terms of their resilience and speed. 

\section*{Acknowledgments}
We thank Marco Schiano and Andrea Di Giglio for their contribution to collecting the dataset used in this work. We gratefully acknowledge Infinera (now part of Nokia) for providing the Groove G30 transponder.

\bibliographystyle{IEEEtran}
\bibliography{bibliography}


 





\end{document}